\documentclass{article}
\pdfoutput=1

\usepackage{arxiv}

\usepackage[utf8]{inputenc} 
\usepackage[T1]{fontenc}    
\usepackage{hyperref}       
\usepackage{url}            
\usepackage{booktabs}       
\usepackage{amsfonts}       
\usepackage{amsmath}
\usepackage{nicefrac}       
\usepackage{microtype}      
\usepackage{cleveref}       
\usepackage{lipsum}         
\usepackage{graphicx}
\usepackage{natbib}
\usepackage{doi}

\usepackage{placeins}
\usepackage{upgreek}
\usepackage{multirow}
\usepackage{setspace}

\title{How Participants Respond to Computer Delays}

\usepackage{authblk}

\newbox{\orcid}\sbox{\orcid}{\includegraphics[scale=0.06]{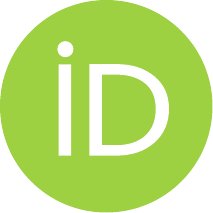}} 
\author[1]{%
	D\'{e}sir\'{e}e~Scholz%
}
\author[2]{%
	Linda~Graefe\thanks{\texttt{Linda.Graefe@uni-jena.de}}%
}
\author[3]{%
	\href{https://orcid.org/0000-0001-9602-3482}{\usebox{\orcid}\hspace{1mm}Thomas M.~Prinz\thanks{\texttt{Thomas.Prinz@uni-jena.de}}}%
}
\affil[1]{Friedrich Schiller University Jena, Jena, Germany}
\affil[2]{Department of Methodology and Evaluation Research, Friedrich Schiller University Jena, Jena, Germany}
\affil[3]{Course Evaluation Service, Friedrich Schiller University Jena, Jena, Germany}

\hypersetup{
pdfauthor={Desiree Scholz, Linda Graefe, Thomas M. Prinz},
pdftitle={How Participants Respond to Computer Delays}
}
\begin{document}
\maketitle

\begin{abstract}
Reaction time studies with computers investigate how and how quickly participants respond to changing sensory input. They promise simple and precise measurement of time and inputs and offer interesting insights into human behavior. However, several previous studies have discovered imprecisions in timing appearing as delays, depending on the browser, software and programming used for conducting such studies. Since the accuaracy of the collected data is widely discussed, we aim to provide new results on the effect of unintended delays on participants' behavior. For this purpose, a new reaction time study was conducted. Computer delays were added to the experiment to investigate their effects on participants' performance and repulsion. Minimal changes in participants' behavior did occur and should be furtherly investigated, as the power of this study was rather low and might not have uncovered all underlying effects. The following report details our study design and results and offers several suggestions for improvements in further studies. 
\end{abstract}

\section{Introduction}\label{Introduction}
Reaction time studies investigate how and how quickly participants react on changing sensory inputs \citep{LatencySimpleReactionTime}. Especially in human-computer interaction (HCI) and psychology, researchers use computers for their experiments since they promise cheap, accurate, and precise measurements of time and inputs \citep{HistoryComputerScience,Experimenter}. However, the greatest possible accuracy would require instanteneous hardware and software responses, which is almost never possible, as proven in numerous recent studies with differing hardware/software \citep{AccuracyOnlineExperiments, TimingMegaStudy, MethodIssuesRT, AccuracyMac}. Of course, no researcher wants to conduct an unintended non-objective experiment, i.\,e., that unnoticed variations in the experiment's computer system (such as delays) influence the participants' behavior and, therefore, the experimental results. 

Although non-objective experiments are a topic of research, delays in a computer's response are usually considered random and, therefore, negligible. This assumption is supported by the replication of general results, including effect sizes, when comparing web- and lab-based research \citep{UIPerformanceEvaluation, LabVsWeb, ReportDropouts, JavaScriptPsychophysics}. In contrast to these reports, \cite{LatencySimpleReactionTime} and \cite{MethodIssuesRT} discovered discrepancies in the results of previous reaction time studies and a tendency towards increasing reaction times over the course of several decades. 
The impact of mechanical inaccuracy is usually investigated by comparing results of different devices and / or approaches. One can also investigate response patterns after artificially introduced delays in order to say whether they directly influence participants' behavior, as done by \cite{SRTProblemSolvingBehavior}, \cite{SRTUserResponseTime}, \cite{SRTMethodOfPay} or  \cite{EffectsOfDelay}. However, the results of these studies vary: They show either large or no effects. The reasons for these differences currently are a matter of speculation.

Their research emphasizes the need for further investigations on this topic, as ignoring unintended delays could systematically affect participants' experimental results and, therefore, the validity of the study.

Systematic differences in results can have strong impacts. For example, there are reaction time tests in the medical context, where deviations from reaction times of healthy patients provide diagnostic information. This diagnostic information may provide access to help, interventions, and treatments.

False negative results can, therefore, potentially lead to long-term deficiencies. In HCI research, reaction times are important to measure how and how quickly users interact with computers. Hence, unintentional computer delays in experiments could lead to misinterpretations of HCI phenomena. For example, suppose that a prompt is displayed to a user in a software after a predefined time when the user does not have interacted with the computer. If delayed system responses prevent users from interacting with the software, the prompt will be displayed incorrectly. In the worst case, reactions of users to the software are interpreted as responses to the prompt and, therefore, lead to unintended effects on the software. Such unintended reactions can have strong effects, as \cite{InteractionInterferences} show.

This paper presents the results of a new reaction time study that investigates the effects of computer delays on participants' performance and repulsion. The study used a simple experiment in which participants had to decide which of two numbers was greater. Participants were divided into four groups: a control group without additional delays and three experimental groups with different variability of delays. We hypothesized that higher variability in computer delays would lead to higher reaction times and repulsion among participants. We also hypothesized that error rates would increase with additional delays, especially when variability is high. A total of $100$ persons participated in the study. 

We were unable to confirm our main hypotheses. Additional explorative analyses show minimal changes in participants' overall response speed as well as an apparent impact of age and gender in the behavior following the delays. 

The paper is structured as follows: Section ~\ref{sec:RelatedWork} provides an introduction to the human-computer interaction process. Subsequently, it describes our newly conducted experiment with our hypotheses in Section~\ref{sec:Experiment}. Section~\ref{sec:Analysis} details the analyses used and the results of the study. These results are further discussed in Section~\ref{sec:Discussion}. Finally, a conclusion is drawn in Section~\ref{sec:Conclusion}.

\section{Theoretical Background}
\label{sec:RelatedWork}

Previous research describe the general phenomenon of interruptions in information technology.  \cite{InformationTechnologyInteruption} distinguish interruptions between \emph{intrusions} and \emph{interventions}. Both are caused by external factors (e.\,g., pop-ups of incoming messages). Intrusions do not belong to the current primary task. Therefore, they do not lead to a long-term change in behavior in relation to the task at hand. Interventions, on the other hand, show deviations between the expected and the actual performance of the primary task. As a result, they lead to long-term changes in behavior to avoid further interventions. \cite{InformationTechnologyInteruption} further make distinctions between different types of intrusions: Some are \emph{informational} but irrelevant in the current situation (e.\,g., at work), others require \emph{actions} that have nothing to do with the primary task but are relevant to the general situation, and others are caused by the computer (\emph{system} intrusions). In a controlled laboratory environment, as it often used in psychological or HCI-related experiments, informational and action-related intrusions are usually an intended part of the experiment. System intrusions are usually unintended. In the present work, delayed responses (system intrusions) are intentionally added in a reaction time study to investigate their effects on participants.

Reaction time studies, as the name suggests, examine how and how quickly participants respond to changing input signals. The time needed to react to the changing input signal is generally referred to as \emph{reaction time}. \cite{SRTProblemSolvingBehavior} propose a distinction in user response time that is necessary to interpret it correctly in both natural and experimental settings. They argue furthermore that the \emph{human-computer interaction process} (HCIP) consists of two primary time-spans: The \emph{user response time} and the \emph{system response time}.

\begin{figure}[tb] 
	\centering		
	\includegraphics[width=0.8\textwidth]{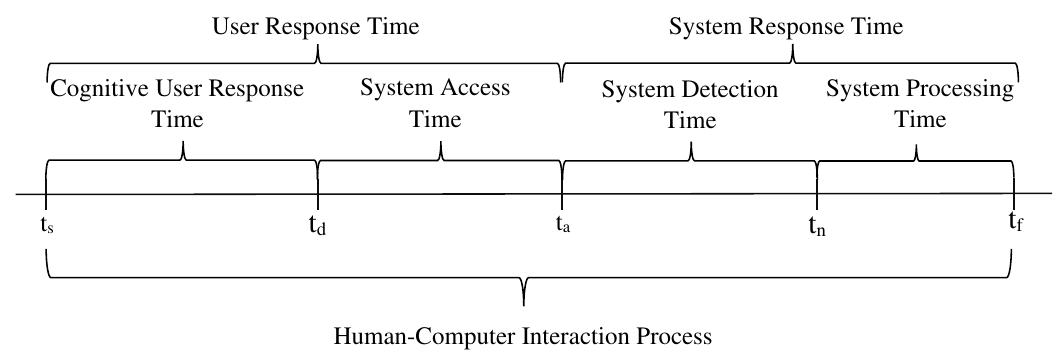}
	  \caption{Response times in human-computer interaction processes}
\label{fig:VisualizationHCI}
\end{figure} 

Figure~\ref{fig:VisualizationHCI} illustrates both 
time spans. The user response time comprises the time from the beginning of a measurement $t_s$ ($s$ for \textit{start}) to the point in time $t_a$ ($a$ for \textit{action}) when the desired action (e.\,g., clicking a button) is completed. It can be divided into the \emph{cognitive} user response time, which starts at $t_s$ and lasts until the point in time $t_d$ ($d$ for \textit{decision}) when the user has made a decision about action to take, and the immediately following \emph{system access time}, which comprises the performed user action to be finished.
The second time span of the HCIP, the system response time, includes the time after the user action is completed $t_a$ until the point in time $t_f$ ($f$ for \textit{finish}) when the result of the action is presented (e.\,g., displaying a new text on a screen). While describing their precautions to measure all possibly occuring delays \cite{LatencySimpleReactionTime} further divide the system response time into two time spans. We have included them in the model as follows: 
First \emph{system detection time}, which describes the time it takes for the computer to recognize the user's action. The remaining period between the point in time $t_n$ ($n$ for \textit{noticed}), when the computer recognized the action, and the time $t_f$, when the computer has processed effects of the action, is the \emph{system processing time}.

\FloatBarrier

\section{Methods}
\label{sec:Experiment}

Due to the differences in the results of previous research efforts, we conducted a study to gain new insights into delayed system responses in reaction time studies. A main requirement of the experiment was to make the task comparable to simple reaction time experiments from psychology and HCI using a one-step cognitive decision and a two-choice response. As the experiment was used in an online study due to Covid-19, it was ensured that the experiment offered little opportunity for confusion or cheating by using the guidelines suggested by \cite{DosAndDonts} (e.\,g., randomization of data, one-way passwords, etc.). The experiment was tested for conciseness on a group of five volunteers and their comments were integrated into the final experiment. After a second pretest with further eight participants, a high error rate was noticed caused by a simple, remaining misunderstanding in the experimental instructions. This was solved in the final experiment.

\subsection{Experimental Setup}

\begin{figure}[tb]
\centering\includegraphics[width=1.0\textwidth]{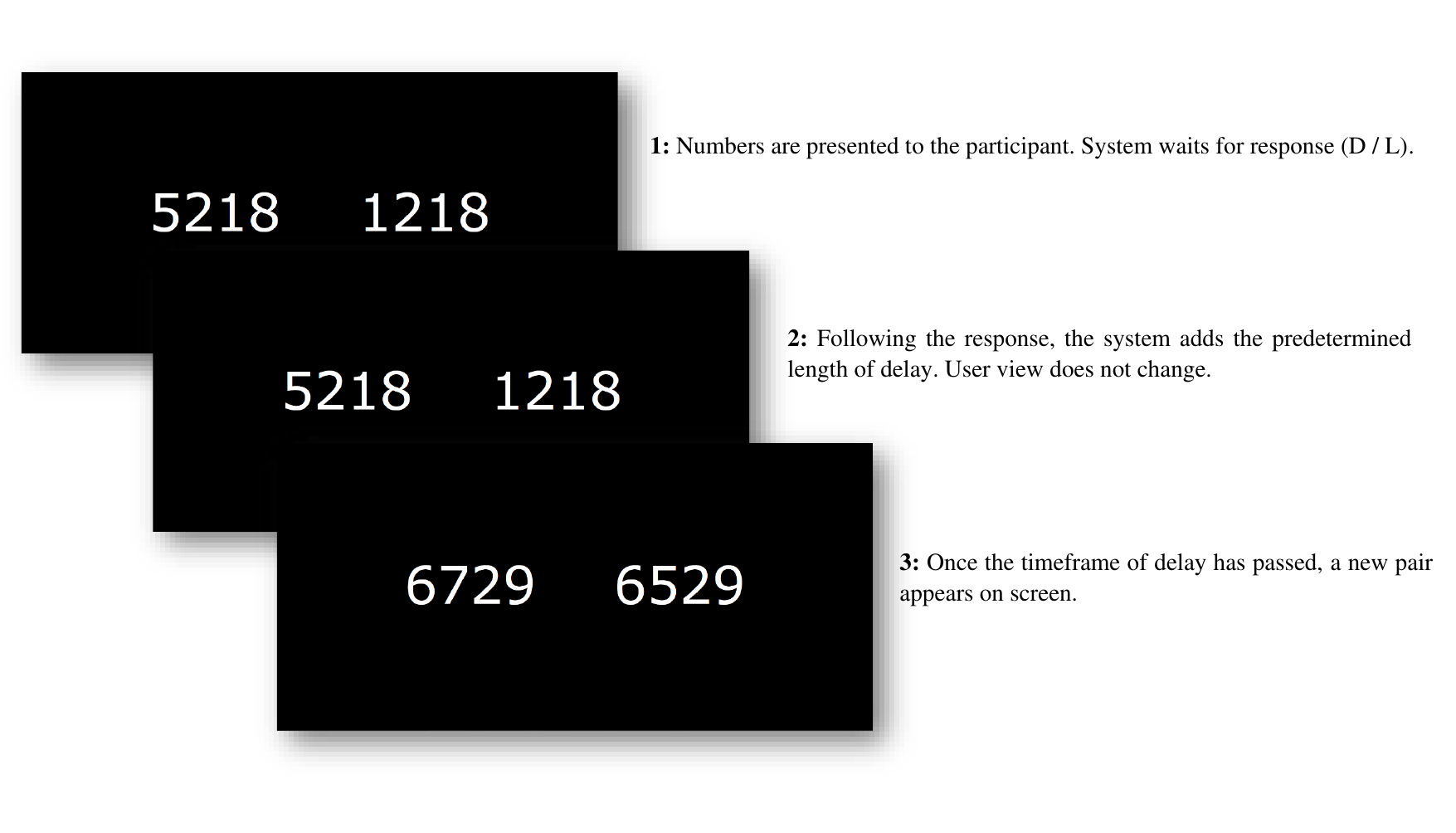} 
\caption{Example Trial.}
\label{fig:ExperimentSetting}
\end{figure}

After all pretests, the final experiment had the following setting: Each participant got a set of $320$ number pairs. Each number has four digits ($1000$ to $9999$) and differs in exactly one digit from the other number in the pair (e.\,g., $1\mathbf{2}34$ and $1\mathbf{4}34$). The choice of numbers of the same size with only one difference in digit is justified to avoid longer reaction times caused by a higher complexity of the comparison \citep{NumberDifferences}. Both numbers were displayed side by side in the internet browser (cf. Figure~\ref{fig:ExperimentSetting}). The participants had to decide which of the two numbers was the greatest. If the left number is greater than the right one, the participants had to use the \emph{D} key, otherwise \emph{L}.

The entire experiment was divided into $7$ blocks; two blocks for practice followed by five blocks for the experiment. The two practice blocks comprised $10$ number pairs each, which were the same for all participants. The five experimental blocks comprised $60$ pairs of numbers each. However, each participant got all $300$ pairs in the experimental blocks in a randomized order to avoid position effects \citep{PositionEffect}.

\FloatBarrier

Three different experimental groups were designed to examine different kinds of system delays. Due to the lack of data on average delay durations, we based the mean duration on 1650ms, as suggested by \cite{EffectsOfDelay}.
Participants of Group $1$ got a constant delay of $1650$ ms for each pair of numbers (including the practice blocks); Group $2$ got a delay with low variability ($\sigma = 0.5s$) and a mean of $1650$ ms; finally, participants of Group $3$ got a delay with high variability ($\sigma = 1.5s$) and the same mean of $1650$ ms. In addition to the three treatment groups, a fourth control group (referred to as Group $0$) showed no intended delays.

The delays for Group $2$ and $3$ were calculated using $\gamma$-distributions with the mean values and standard deviations given, in line with \cite{SRTProblemSolvingBehavior}. The choice was made based on the unaffectedness by the lack of negative values as well as the right-skew, since shorter delays are more common than long delays \citep{SRTUserResponseTime}.
 The distributions result in delays from $562.1$ to $3206.9$ ms for Group $2$ and from $17.3$ to $7697.0$ ms for Group $3$. The delays were calculated once and were the same $320$ delays for all pairs of numbers (but in a random order) within each experimental group.
 
Figure~\ref{fig:ExperimentSetting} demonstrates the sequence of events for one trial.

\subsection{Measured Times and Data}

The experimental set-up measured three different time intervals:%
\begin{description}
	\item[System time $\Delta t_{sys}$] describes the time between taking a new pair of numbers, waiting for the intended delay, and displaying the pair in the browser for the participants. In the HCIP model previously introduced, this can be interpreted as the system processing time.
	\item[User response time $\Delta t_{user}$] comprises the time interval between displaying the current pair of numbers and pressing either key \emph{D} or \emph{L}. This includes the entire user response time and the system detection time, as explained in the HCIP model.
	\item[Total experiment time $\Delta t_{total}$] starts with the system time measure and ends with the user response time measure.
\end{description}%
The system time $\Delta t_{sys}$ was measured to detect unintentional delays beside the intended delays. That is, if $\Delta t_{delay}$ describes the current delay for the number pair, then $\Delta t_{sys} - \Delta t_{delay}$ is the unintentional delay. To avoid unintentional delays, the complete experiment was loaded at once as a JavaScript/ECMAScript file such that no reloading of sources over the web was required. The total experiment time $\Delta t_{total}$ was measured to check for further unintentional delays in the time measurements. A more accurate measurement of user response time without the system detection time was not possible due to the necessity to have a user input (key press \emph{D} or \emph{L}) to perform a system action.

Besides the measured times, the participants also had to state their age and gender, as both could be a cause of prolonged reaction times independent of the experimental condition \citep{AgeAndReactionTime,AgeDifferences2,CognitiveDifferenceGender2,CognitiveDifferenceGender}. In addition, after each block (practice as well as experimental), a question was asked about the participants' repulsion, which was rated on a 6-point Likert scale. At the end, there was again a question on repulsion in relation to the whole experiment.

Sometimes, participants tend to find the experimental hypothesis during the study that influences their behavior \citep{VEE}. Therefore, a question regarding the suspected hypothesis/question was added to the study. A control analysis was conducted that excluded all participants who correctly assumed the experimental manipulation.

\subsection{Hypotheses}
\label{subsec:Hypotheses}

Taking into account the experimental conditions, the following assumptions were made:%
\begin{description}
	\item[Hypothesis 1] An higher variability in system response time is linked to an higher user response time.
	\item[Hypothesis 2] Error rates are larger in conditions with delays in comparison to the control group.	
	\item[Hypothesis 3] The higher the variability in system response time, the higher the repulsion.	
\end{description}

\section{Analyses/Results}
\label{sec:Analysis}

Participants were invited to take part in the experiment described in the previous section. To avoid any effects of previous knowledge about the goal of the study, participants were invited with a cover story to \emph{investigate the effects of age and gender on number perception}. The actual aim of the study was not presented until the individual participants completed their experiments. They had a chance to withdraw their consent to participation at any point during the experiment. All participants were recruited via various university mailing lists in Germany, the website of the German psychology magazine \emph{Psychologie Heute}, and personal requests to friends and colleagues unfamiliar with the project. 

The results are divided into the following parts: First, the composition of the groups and some general results are presented. Afterwards, the influences of the covariates on the measures are described. Subsequently, the results test our hypotheses and provide some alternative exploratory analyses. At the end, the effect on the results of participants' assumptions about the study hypotheses are considered.

\subsection{Composition of Groups and General Results}
\label{subsec:Composition}

The general results and composition of the groups differ in results about participants, distribution of system response times, performance, and repulsion of the participants.

\subsubsection{Participants}

At the end of recruitment, $141$ people took part in the study. One of them had to be excluded because of the age\footnote{According to 
law in the country where the study was conducted, the data of minors may only be used for studies if their custodians have given their explicit agreement.}. Of the remaining $140$ participants, only $100$ actually completed the entire study. They were randomly assigned to the four groups, but the final group sizes were unbalanced ($N_{X = 0} = 29$, $N_{X = 1} = 27$, $N_{X = 2} = 20$, $N_{X = 3} = 24$). Implications of the unbalanced group sizes and the dropout of $40$ participants are discussed later in Section~\ref{sec:Discussion}.

Most participants were female ($N = 66$), only a small minority chose the gender option \emph{diverse} ($N = 5$). Since there were no diversified participants in Group $3$ and the overall data situation is insufficient \citep{Nonbinary}, gender-specific differences can only be interpreted between male and female participants. The groups did not significantly differ in the distribution of these two genders ($\chi^2(3, N = 95) = 1.09$, $p = 0.78$).

The age of the participants ranged from $19$ to $64$ years ($M = 30.32$ years, $SD = 10.45$ years), with no significant differences in mean between the groups ($F(3, 96) = 0.361$, $p = 0.781$).

\subsubsection{Distribution of System Response Times}

We calculated unintended system response times by reducing the measured system time $\Delta t_{sys}$ by the intentionally added delays. This resulted in 320 measurements of unintended system response time per person, which allowed to identify its distribution for each participant. 
Over all participants, the mean system response time varied from $0.23$ ms up to $11.63$ ms ($M = 5.5$ ms, $SD = 4.07$ ms), while the standard deviation ranged from $0.05$ ms to $9.26$ ms ($M = 3.05$ ms, $SD = 2.36$ ms). There is a significant difference in mean system response time across groups ($F(3, 38) = 108.6$, $p < 0.001$) and standard deviation of system response time across groups ($F(3, 39) = 80.343$, $p < 0.001$). Although participants probably did not notice these short intervals, possible side effects are nevertheless discussed in Section~\ref{sec:Discussion}.

\subsubsection{Overall Performance of Participants} 

The average user response time per trial (pair of numbers) of a single participant was between $1224$ ms and $1509$ ms ($M = 1368$ ms, $SD = 54.34$ ms). The values were calculated using the median to avoid being influenced by individual outliers. Although a comparison with other studies is difficult due to differences in the task specification, the research assumes that these are realistic distributions for a mathematical task \citep{DataAnalysis, SpeedReaction}.

The errors made by participants were minimal, with $50$\% of the participants making less than $9$ mistakes ($M = 13.71$, $SD = 18.28$). The lowest number of errors was $1$, the highest was recorded with $125$ errors.

\subsubsection{Repulsion Scores}

Aggregated repulsion across all blocks ranged from $0$ to $4.8$, where $0$ means \emph{no repulsion} and $5$ means \emph{very high repulsion} ($M = 1.3$, $SD = 1.28$). In other words, some participants experienced no repulsion at all, while no participant experienced very high repulsion constantly.

\subsection{Influence of Covariates on Measures}
\label{sec:InfluenceCovariates}

The study measured two covariates, \emph{age} and \emph{gender} of participants. Their influence on measurements is described below.

\subsubsection{Age} 

Using a correlative approach, age was found to significantly influence both reaction time and error rate. With increasing age, participants appeared to become slower, indicated by $r(98) = .27$, $p = 0.006$. This is consistent with previous findings \citep{AgeAndReactionTime, AgeDifferences, AgeDifferences2, LatencySimpleReactionTime}. However, the correlation between age and error rate was found to be negative with $r(98) = -0.23$, $p = 0.022$, indicating either the absence of cognitive decline with increasing age or an increase in diligence. The latter is further supported by a significant negative correlation between error rates and reaction times with $r(98) = -.40$, $p < 0.001$. A linear regression model using both age and error rate as additive factors explains 19.6\% of overall variance in user response time ($NER$ is the number of errors): %
\begin{equation}
\label{eq:ReactionTime-Age+Error}
E(\Delta t_{user}|Age, NER) = 1390.212 - 6.313 \cdot NER + 5.805 \cdot Age   
\end{equation}%
The effects of age on performance were limited, as the correlation between age and repulsion was not significant ($p = 0.339$).

\subsubsection{Gender} 

As already mentioned, the following results only compare male and female participants. 

No significant effects on performance in reaction time ($t(61) = -0.988$, $p = 0.327$) or error rates ($t(38) = 1.056$, $p = 0.298$) were discovered. As this task is a more cognitive one, these results are consistent with the absence of gender differences in intelligence \citep{Intelligence}.

A t-Test was used to examine the effects of gender on repulsion and yielded significant results ($t(61) = -2.5176$, $p = 0.015$). Females apparently rated their repulsion higher than male participants, with a mean difference of $0.65$. These findings also complement previous research on differences in the perception and expression of emotions \citep{AversionGender, AversionGender2}.

\subsection{Effects of Delay Manipulation}
\label{subsec:Effects}

Several tests were carried to investigate the hypotheses introduced in Section~\ref{subsec:Hypotheses}.

\subsubsection{Hypothesis 1}

\emph{A higher variability in system response time is linked to an higher user response time.} An one-way t-test was used to examine whether the mean user response time of Group $3$ was significantly higher than the mean user response time of Group $2$. In addition, since the homogeneity of variance was verified, the t-test is considered robust to non-normality and unequal group sizes according to the guidelines for empirical sciences \citep{TTest}. The analysis yielded a non-significant result $t(40) = -0.468$, $p = 0.321$. \emph{Therefore, the variability of the system response time cannot be considered as an influencing factor on the user response time in this study.}

\subsubsection{Hypothesis 2}

\emph{Error rates are larger in conditions with delays in comparison to the control group.} Since this hypothesis suggests a difference in means of all experimental groups in comparison to the control group, a one-factorial Anova was used to investigate the differences in error rates. Homogeneity of variance was verified so that unbalanced group sizes were of little importance \citep{AnovaUnbalanced}. The result was not significant ($F(3,96) = 0.258$, $p = 0.855$). \emph{The assumption of an unequal mean error rate between the groups cannot be confirmed.}

\subsubsection{Hypothesis 3}

\emph{The higher the variability in system response time, the higher the repulsion.} In this case, all experimental groups were compared pairwise with another one-factorial Anova, assuming that unbalanced group sizes have no influence due to the homogeneity of variance \citep{AnovaUnbalanced}. The resulting $F(2,68) = 0.450$ and  $p = 0.639$ imply that \emph{no group mean in repulsion differs significantly from another group mean, leading to the rejection of this hypothesis}.

\subsection{Alternative Analyses}

The previous analyses treated the effects of covariates as less important (1) due to the lack of research on the influence of participants' age and gender on perceptions of system response times and (2) due to non-significant differences in the distribution of these variables between groups. Therefore, we exploratively re-examined the impact of system response times per se on trial response time (instead of looking at them group-wise). In a second additional analysis, we investigated the suitability of several complex linear regression models to explain the total time for all experimental trials.

\subsubsection{Trial User and System Response Time} 

All $30{,}000$ experimental measurements of system response times (including manually added delays, i.\,e., $\Updelta t_{sys}$) and user response times ($\Updelta t_{user}$) were used for correlative analysis and linear regression. The correlation appeared to be small ($r(29998) = -.07$) but significant ($p < 0.001$). To avoid bias due to outliers, a second correlative approach was taken, excluding all user response times lower than $291.633$ ms and higher than $2500.613$ ms. Such values could be caused either by accidental, reflexive typing or by (environmental) distractions \citep{DataAnalysis}. $93.3$\% of the data was used for the second analysis, revealing a still significant correlation of $r(28095) = -.13$, $p < 0.001$. Accordingly, the following linear regression model explains about $1.5$\% of the total variance in user response times: %
\begin{equation}
\label{SRTURTEq}
E(\Updelta t_{user}|\Updelta t_{sys}) = 1454 - 0.044 \cdot \Updelta t_{sys}
\end{equation} 
However, the results of this analysis have to be interpreted with caution, since the number of trials is quite large, which could influence the reports of significance, but derived from repeated measurements on the 100 participants.

\subsubsection{Complex Models with Predictors and Interactions}
\label{Alt2}

Five models seemed interesting to compare their ability to describe the data, in particular the time to complete all experimental trials, the number of errors, and repulsion:%
\begin{description}
  \item[Model 1] Group as predictor.
  \item[Model 2] Group and age as predictors (without interactions).
  \item[Model 3] Group, age, and gender as predictors (without interactions).
  \item[Model 4] Group, age, and gender as predictors (interaction between group and age).
  \item[Model 5] Group, age, and gender as predictors (interactions between all variables).
\end{description}%

For the comparison of these (multiple) linear regression models, the maximum-likelihood-ratio-test was used, and all models were specified as reference group models, with \emph{control} as the referenced group. This gives rise to multiple indicator variables $I_{group = x}$ that amount $1$ whenever $group = x$ and $0$ in all other cases.

	\paragraph{Overall time to complete the experiment} Comparing all five models in terms of the user response time, Model 2 turned out to be most suitable at describing the user's time to complete the experiment ($\chi^2(1, N = 95) = 8.239$, $p = 0.004$). Three parameters differ significantly from zero with an $\alpha$-level of $.05$. The model below explains $13$\% of the total variance with $^*$ identifying significant coefficients: %
\begin{equation}\label{EqKomplex1}
\begin{aligned} 
E(\Updelta t_{user}|\Updelta t_{sys}, Age) = & 402196.6^* -49387.5 \cdot I_{group=1}\; - \\
& 53977.8 \cdot I_{group=2} - 53729.6^* \cdot I_{group=3} \; + \\
& 2617.6^* \cdot Age
\end{aligned}
\end{equation}%

\FloatBarrier
\paragraph{Number of errors (NER)} Both Model 2 ($\chi^2(1, N = 95) = 4.932$, $p = 0.026$) and Model 5 ($\chi^2(7, N = 95) = 40.455$, $p < 0.001$) are comparatively more suitable than the baseline model. Comparing these two models individually, Model 5 fits even better ($\chi^2(11, N = 95) = 43.443$, $p < 0.001$). However, only four of the model's parameters are significant, as shown in Table~\ref{tab:EqKomplex2}. Overall, Model 5 explains $41$\% of the variance in the number of errors (NER) between participants.

\begin{table}[tb]
\caption{Significant regression coefficients for describing the number of errors.}
\label{tab:EqKomplex2}
\centering
\begin{tabular}{llrrr}
	\toprule
	\multicolumn{5}{c}{Regression coefficients} \\
	\hline
	Description & & Coefficient & t-Value & p-Value \\
	\hline
Parameter    & $I_{group=1}$                              &  186.6 &  5.093 & $<$ 0.001 \\
Interaction & $I_{group=1}$ and Age                    &   -5.3 & -4.622 & $<$ 0.001 \\
Interaction & $I_{group=1}$ and $I_{gender = female}$ &   -214.1 & -5.330 & $<$ 0.001 \\
Interaction & $I_{group=1}$ and Age and $I_{gender = female}$ & 6.0 & 4.757 &  $<$ 0.001 \\ 
	\bottomrule
\end{tabular}
\end{table}

\paragraph{Repulsion} 
For the relationship between repulsion and all measured variables, Model 3 seems to be most suitable ($\chi^2(1, N = 95) = 7.220$, $p = 0.007$). The four significant parameters of this model, explaining $13$\% of the variance in repulsion, are described in Table~\ref{tab:EqKomplex3}.

\begin{table}[tb]
\caption{Regression coefficients in a model describing repulsion.}
\label{tab:EqKomplex3}
\centering
\begin{tabular}{llrrr}
	\toprule
	\multicolumn{5}{c}{Regression Coefficients} \\
	\hline
	Description & & Coefficient & t-Value & p-Value \\
	\hline
	Intercept & $group = control$    &  1.50  &  3.368 & $<$ 0.001 \\
	Parameter & $I_{group = 2}$        & -0.88  & -2.495 &   0.014 \\
	Parameter & $I_{gender = female}$  &  0.71  &  2.651 &   0.010 \\
\bottomrule
\end{tabular}
\end{table}

\FloatBarrier

\subsection{Checking Participants' Assumptions of Study Hypotheses}

$79$ participants made assumptions about the research question of the study. The assumptions were finally divided into $7$ categories. The interested reader can find all categories with some examples in Table~\ref{tab:tabHypo}. After repeating the analyses described in the Sections~\ref{subsec:Composition} to \ref{subsec:Effects}, we replicated the general results in terms of their significance or insignificance. In the case of significant results, the difference was marginal.

\begin{table}[tb]
\caption{Categorized guesses of participants on possible research questions. The examples are translated 
into English. The category \emph{Critical} describes all hypotheses that require an understanding of the experimental manipulation. If participants made more than one suggestion, the most detailed (or critical) one was preferred.}
\label{tab:tabHypo}
\centering
\begin{tabular}{p{3.5cm}cp{10.5cm}}
	\toprule
	Category & N & Examples \\
	\hline
	Concentration \& Reaction & 14 & (1) Difficulties concentrating on a task. \newline (2) Monotony and Perception. \\
	\hline
	Comparing numbers & 13 & (1) Which digit causes most errors? \newline (2) Can digit switches influence the result? \\
	\hline
	Errors $\leftrightarrow$ Repulsion & 12 & (1) Higher error rate causes higher repulsion. \newline (2) Relationship between error rate \& aversion. \\
	\hline
	Critical & 11 & (1) Higher repulsion due to longer delays. \newline (2) Patience \& Attention. \\
	\hline
	Resistance to stress & 10 & (1) Comparing numbers during time pressure. \newline (2) How stressful is flow disruption? \\
	\hline
	Duration $\leftrightarrow$ Performance & 6 & (1) Higher errors with longer duration of the experiment. \newline (2) Higher errors after comparing many numbers. \\
	\hline
	Other &  13 & (1) Repulsion and repetitive tasks. \newline (2) Younger participants have better performance for number comparison. \\
	\bottomrule
\end{tabular}
\end{table}

\section{Discussion}
\label{sec:Discussion}

In the following, the above-mentioned results of the study are discussed. The discussion is divided into the following parts: First, the effects of variations in system response times are detailed. This is followed by a discussion about the reductions in user response times over the duration of the experiment. Next, the discussion considers the impact of dropouts (break-offs) and provide some analyses of key presses as an alternative indicator of repulsion. Finally, the limitations of the present work are discussed.

\subsection{Effects of Variation of System Response Times}

Since none of our analyses led to significant results, one could assume that manipulating the system response time has no impact on the participants. However, when age and gender are included in a multiple regression model, a significant influence of the experimental group (mainly Group $1$ and Group $2$) is becoming apparent. Interestingly, this influence was \emph{positive}: increasing the system response time led to a faster completion of the experiment by the participants, almost one minute shorter than in the control group without additional delays.

Group 1 with constant delays of $1650$ ms had several significant interaction effects with age and female gender on participants' individual error rate. As previous studies do not detail the effects of combined covariates, this is a new finding that should be further investigated. Another interesting finding is the lack of significant repulsive responses for all groups except Group $2$.

Although these findings were surprising and contradict our hypothesis, there are plausible explanations for the results based on two different models of human-computer interaction: %
\begin{description}
	\item[Processing Time:] In the interaction between humans and computers, both partners are limited in their processing speed. As a result, a computer's response should be fast but slow enough for the human to actually process the information presented. Some limitations lie in the memory capacity of the human \citep{DisplayRate}: Although working memory can store information for up to $30$ seconds \citep{MemoryCapacity}, interferences with previously presented information are possible \citep{MemoryInterference, OverwritingMemory}.	
	
	With regard to the present study, the choice of the greater number might be difficult due to the similar characteristics of the trials (two four-digit numbers), as the old pair is still in mind. Therefore, adding delays grants participants more time to lose focus on the old pair in memory. In part, this approach is also supported by the significant negative correlation between user response time and the number of errors. However, reflexive key presses or general distractions from the environment cannot be excluded as a cause for this correlation \citep{DataAnalysis}.
	
	\item[Perceiving a natural interaction:] \cite{RTConversationalTransactions} argued that an interaction between humans and computers is similar to an interaction between people, where only one partner has an active goal. In our case, the participants' main goal was to complete the experiment, with sub-target goals being the completion of each trial \citep{HCIOverall}. \cite{RTConversationalTransactions} described that the perfect speed of the interaction process between humans and computers depends on the nature of the goal/subtype of interaction. Either way, the process should ideally be faster than four seconds, preferably two. However, in some cases (e.\,g., when a new task is requested), it can take up to fifteen seconds without this being perceived as negative. Miller emphasizes that all times are based on his experience as behavioral psychologist and there is no explanation of what would happen if the computer react almost instantaneous. \cite{WebsiteDelays} have shown that the optimal system reaction time is at $0.5$ seconds and, therefore, \emph{not} instantaneous.
\end{description}

\subsection{Decreased Response Time over the Experiment's Duration}

Previous studies report an increase in user response times with longer experimental duration \citep{MentalFatigue}, which made us explore why our study showed a different result.  

The simplest explanation is that our experiment was simply too short to induce fatigue effects (participants were able to complete the study in less than half an hour if they decided not to take any of the breaks offered). Therefore, the trend towards faster user response time could be due to the training \citep{PracticeReaction}. This explanation is further supported by the loss of a significant linear trend when twenty practice trials are included in the analyses, which familiarized participants with the task.  

Another investigation concerned the effects of faster response times due to practice, in particular, whether this effect is influenced by the system response time and its variability. This was investigated by comparing the correlations of the different groups. There was no significant trend in the median user response time ($p_1 = 0.494$ und $p_2 = 0.062$) for either Group $1$ or Group $2$. However, in Group $0$ and Group $3$, the trend towards faster response times ($r(298) = -.12$, $p = 0.033$ and $r(298) = -.17$, $p = 0.003$, respectively) was rediscovered. It seems that the trend of faster response time correlates with the system response time when there is no such response time or it has a high variability. However, this is a singular result. Further research is needed to replicate these findings and investigate their causes.

\subsection{Dropout (Break-offs)}

Online studies are usually and unfortunately coupled with a dropout (break-off) of participants. Although we tried to reduce the potential that could lead to dropouts \citep{DropoutInternet} (e.\,g., by asking personal information first), we still had a high dropout rate. In the following, we discuss the impact of dropouts (as suggested for example by \cite{ReportDropouts}).

Each participant who did not complete the study had at least some experience with the experiment. Among the different groups, $3$ participants left the control group, $8$ Group $1$, $13$ Group $2$, and $12$ Group $3$. Four participants answered all pairs of numbers but not the final overall repulsion question.

The reasons for participants to leave the study cannot be determined. Time and environment could be important factors \citep{CauseDropout}. Since the number of dropouts is higher in the experimental groups than in the control group, the repulsion could be too high due to the long system response times, so that participants left the study in frustration. If some participants left the study frustrated, this could support hypothesis 3. However, there is no additional evidence or data to support this assumption. Since hypothesis 3 was not supported by the study results, it seems unlikely that the frustration was high enough that participants left the study. 

Some of the participants who left the study may have participated (and completed) the study again at a later time. This could have influenced their behavior and, because of differences in system response times between both participations, could have betrayed the goal of the study. However, it is generally difficult to avoid double participation in anonymous online studies. Future studies could include a \emph{Continue later} mechanism, but this could also influence the results if there is a large break in the experiment.

\subsection{Key Presses as Indicator of Repulsion}

Participants had to use either key \emph{D} or \emph{L} to answer which of the two presented numbers was the greatest. Once they had given the answer, it was not possible for the participants to influence the duration of the delay. However, it could be possible that some participants became impatient due to the long system response times and used some keys to check whether the answer was registered by the system or the system itself needed another input to present the next pair of numbers. Thus, the keys that were pressed impatiently between the answer for one pair of numbers and the presentation of the next pair were blocked. Whether and how much keys were pressed could be an alternative indicator for repulsion.

\begin{figure}[tb]
\begin{center}
\includegraphics[width=0.6\textwidth]{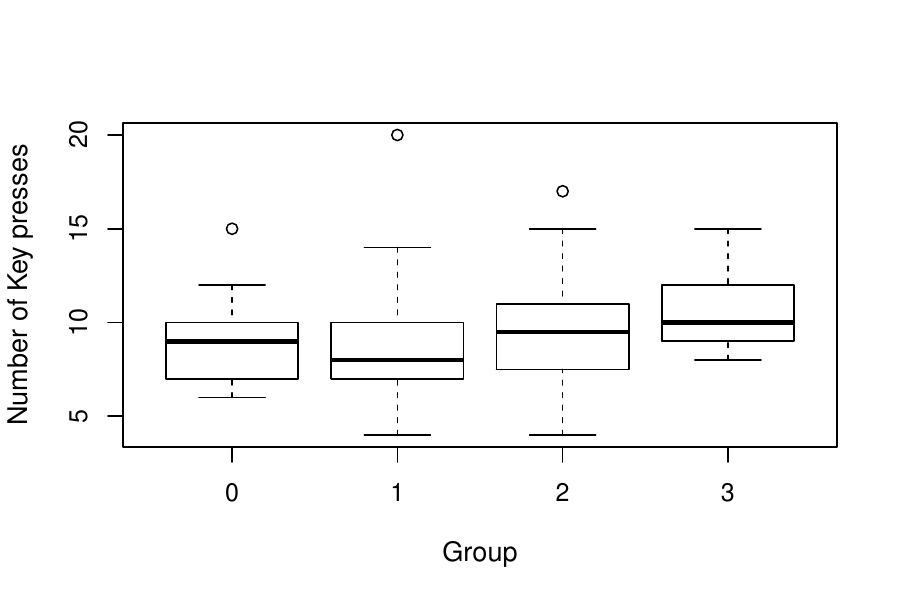}
\caption{Key presses in dependence of the experimental group.}\label{Repulsion}
\end{center}
\end{figure}

Impatiently pressed keys can be separated into two groups: \emph{useful}, i.\,e., \emph{D} and \emph{L} to give an extra response attempt, and \emph{non-useful} keys (all others). Useful keys were used between $3$ and $19$ times per participant ($M = 8.96, SD = 2.69$), while non-useful keys were pressed between $0$ and $3$ times ($M = 0.47, SD = 0.717$). Figure~\ref{Repulsion} takes a closer look at the amount of useful and non-useful key presses between the different experimental groups. In addition, an one-factorial Anova was used to compare all group means. Since Levene's test for homogeneity of variance was not significant ($F(3,96) = 1.843, p = 0.145$), we assumed that there were no influences from unbalanced group sizes \citep{AnovaUnbalanced}. The Anova itself also yielded no significant results ($F(3,96) = 1.811$, $p = 0.150$). As a conclusion, the key presses do not significantly differ between the groups. This is also supported by Figure~\ref{Repulsion}. In other words: if impatient key presses are an alternative indicator of repulsion, then repulsion did not differ significantly between groups. In this study, there was no correlation between the self-reported repulsion score and the number of impatient key presses. 

\subsection{Limitations and suggestions for further studies}

Although this study led to interesting results, it has some limitations that are explained in detail below.

\subsubsection{Sample Sizes and Task Choice} 

The task was explicitly designed for this study by taking into account the time constraints of participants in online studies, typical reaction time tasks, and an easy implementation, measurement, and influence of the delays (due to Covid-19 restrictions, we could not use external measures in a laboratory setting \citep{LatencySimpleReactionTime, RTAccuracy, WebAnimationMethods}). Using a custom task, however, complicates the comparison of results with previous studies and makes it even more complex to determine the pre-hoc effect and the required sample sizes, especially due to the lack of information on expected effect sizes \citep{PowerAnalysis}. Therefore, the minimum sample size of $100$ participants was determined with regard to other studies on this topic and by our own assumptions about what would be possible within the given time limit.

Power calculations help determine sample sizes to find effects of different intensities. To find measurable minimal effects, higher sample sizes are necessary in contrast to finding high effects, where smaller samples are sufficient. Table~\ref{TabEffects} shows different sample sizes for our hypotheses 1-3 and alternative analyses 1 and 2 with models 1 and 3. The sample sizes in the table were calculated using G*Power \citep{GPower1, GPower2}. 

Table~\ref{TabEffects} suggests that our samples allow us to find measurably high effects with an acceptable power ($>$ .80). Confirming of our hypotheses failed as explained. For this reason, our hypotheses do not appear to have a high effect. Previous research suggests that the effect, if present at all, is rather small. Further research on this topic should consider higher sample sizes in order to find smaller effects and to reexamine hypotheses 1-3 against this backdrop. At this stage, we cannot definitively exclude effects of delay. The conclusions of previous studies \citep{EffectsOfDelay, SRTUserResponseTime, SRTMethodOfPay} should in any case be treated with caution, as their sample sizes were also rather small.

The power results of Table~\ref{TabEffects} look more promising regarding our additional analyses. The sample sizes allow acceptable conclusions for even moderate effect sizes with acceptable power above 0.80.

\begin{table}[tb]
\caption{Each section includes six lines. Lines (1) and (2) show the required effect size to achieve a power of .95 / .80 given the underlying sample size. Lines (3) and (4) show the achieved power given a large and medium effect size. Lines (5) and (6) present the required sample size for a power of .95, assuming a medium / small effect size. Alternative Analysis 2 M2 could not adequately be investigated in regards to its power due to the lack of information on power calculation in complex linear models with multiple interactions.}
\centering
\begin{tabular}{crrr}
\toprule
Hypothesis & Sample Size & Effect Size & Power \\
\hline
\multirow{6}{*}{Hypothesis 1, Effect Size $d$} & 20 | 24 & \textbf{1.013} & 0.950 \\
 & 20 | 24 & \textbf{0.765} & 0.800 \\
 & 20 | 24 & 0.800 & \textbf{0.830} \\
 & 20 | 24 & 0.500 & \textbf{0.491} \\
 & \textbf{88 | 88} & 0.500 & 0.950 \\
 & \textbf{542 | 542} & 0.200 & 0.950 \\
\hline
\multirow{6}{*}{Hypothesis 2, Effect Size $f$} & 100 & \textbf{0.423} & 0.950 \\
 & 100 & \textbf{0.330} & 0.800 \\
 & 100 & 0.400 & \textbf{0.920} \\
 & 100 & 0.250 & \textbf{0.518} \\
 & \textbf{280} & 0.250 & 0.950 \\
 & \textbf{1724} & 0.100 & 0.950 \\
\hline
\multirow{6}{*}{Hypothesis 3, Effect Size $f$} & 71 & \textbf{0.477} & 0.950 \\
 & 71 & \textbf{0.377} & 0.800 \\
 & 71 & 0.400 & \textbf{0.848} \\
 & 71 & 0.250 & \textbf{0.438} \\
 & \textbf{252} & 0.250 & 0.950 \\
 & \textbf{1548} & 0.100 & 0.950 \\
\hline
\multirow{6}{*}{Alternative Analysis 1, Effect Size $\rho$} & 28097 & \textbf{0.022} & 0.950 \\
 & 28097 & \textbf{0.017} & 0.800 \\
 & 28097 & 0.500 & \textbf{1.000} \\
 & 28097 & 0.300 & \textbf{1.000} \\
 & \textbf{134} & 0.300 & 0.950 \\
 & \textbf{1289} & 0.100 & 0.950 \\
 \hline
\multirow{6}{*}{Alternative Analysis 2 M1, Effect Size $f^2$}& 95 & \textbf{0.168} & 0.950 \\
 & 95 & \textbf{0.105} & 0.800 \\
 & 95 & 0.350 & \textbf{1.000} \\
 & 95 & 0.150 & \textbf{0.924} \\
 & \textbf{107} & 0.150 & 0.950 \\
 & \textbf{776} & 0.020 & 0.950 \\
 \hline
\multirow{6}{*}{Alternative Analysis 2 M3, Effect Size $f^2$} & 95 & \textbf{0.189} & 0.950 \\
 & 95 & \textbf{0.120} & 0.800 \\
 & 95 & 0.350 & \textbf{1.000} \\
 & 95 & 0.150 & \textbf{0.888} \\
 & \textbf{119} & 0.150 & 0.950 \\
 & \textbf{863} & 0.020 & 0.950 \\
 \bottomrule
\end{tabular} \label{TabEffects}
\end{table}

\subsubsection{Repulsion and Dropout}

If the participants who left the study before completing their tasks did so because of frustration, then this higher repulsion does not appear in our data and, therefore, not in our analyses. As a consequence, the missing data could lead to strong interaction with the results. 

Our simple measurement of repulsion with a single question after each experimental block was used to keep the study short, as it was an online and not a laboratory setting. As a consequence, no pre-existing emotion-theoretical model can be applied (such as in \cite{EffectsOfDelay}) and an accurate measurement of repulsion cannot be guaranteed.

\subsubsection{Differences in Unintended System Response Times}

\begin{figure}[tb]
\begin{minipage}[t]{0.5\textwidth}%
\includegraphics[width=1.0\textwidth]{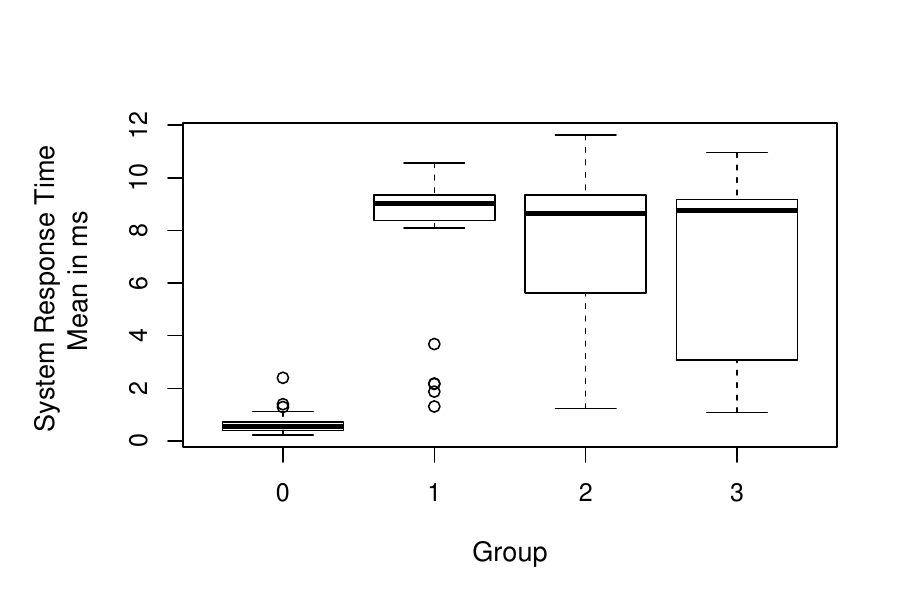}
\end{minipage}%
\begin{minipage}[t]{0.5\textwidth}%
\includegraphics[width=1.0\textwidth]{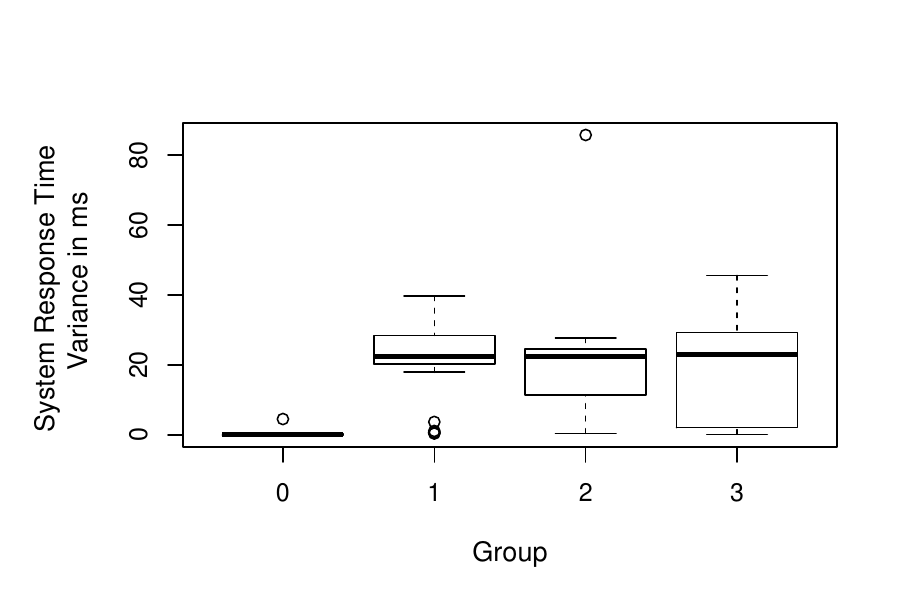}
\end{minipage}
\caption{Distribution of unintended system response times and their variances among groups.} \label{VertSystem}
\end{figure}

As reported in Section~\ref{sec:Analysis}, there were significant differences in unintended system response times between the groups (i.\,e., the delay was not added on purpose). Figure~\ref{VertSystem} illustrates these unintended response times. The largest difference occurred between the control group $0$ and all experimental groups $1$-$3$. Although the experiment was carefully implemented to avoid side-effects, the overhead of adding a delay led to unintended system response times in the experimental groups. Compared to the intended additional delays, the unintended system response times are small. Furthermore, several studies report that the threshold for perceiving visual changes is between $50$ and $90$ Hz ($11$ to $20$ ms, respectively). Most of the values reported here are not within this time-span and seem negligible. If further studies, such as that of  \cite{ImageRates}, indicate that perceiving visual changes is faster in some situations, the results of our study would need to be interpreted further in terms of the unintended system response time. Since the variations are mostly consistent with the experimental manipulation, we decided against regrouping participants based on their distributions of unintended system response times.

\section{Conclusion}
\label{sec:Conclusion}

Unintended system response times in reaction time studies may affect the measurement accuracy and the participants' behavior in the experiment. In the worst case, such system times systematically interfere with the study results and lead to false conclusions. For this reason, this paper presented a new study investigating the relationship between variation in system response time, especially in variability, participant's performance, and repulsion. Our three postulated main hypotheses were not supported by the data: reaction times, repulsion, and error rates appear to be almost independent of computer delays and the height of variability. However, the lack of dectable effects might only be due to the tests comparatively small power.
Our additional analyses show only small positive influences of increased system response times, primarily when including age and gender as additional factors. In simple words: When system response time increases, reaction times and error rates seem to decrease slightly. Two theories have been presented to explain this effect. In general, the results suggest that participant's behavior in reaction time studies is relatively stable when it comes to delays of modern computers. Therefore, such behavior changes do not seem to deliver an explanation for frequently discovered historical trends and replication difficulties \citep{MethodIssuesRT, ReplicationFailureAccuracy, LatencySimpleReactionTime} 

Reaction time studies are popular in HCI. For this reason, the present paper provides new insights into the effects of unintended system response times on study results and user behavior. It supplements the rare palette of studies on these effects and we hope that it will motivate further research on this topic as well as the consideration of system response times in analyses, evaluations, and discussions in future reaction time studies. Although this paper has shown that reaction time studies seem to be robust in the presence of computer delays, there are also some small effects that should be taken into account.

In future studies on this topic, \emph{un}intentional system response times should be stable across all conditional groups. Since humans seem to need some time to switch from one task to the next, the control groups should not have almost no delays. It might be interesting to add a default delay to all groups (control and experimental). Finally, new studies dealing with repulsion should, if possible, try to measure repulsion more accurately and to prevent participants from leaving the experiment. Both seem to be only possible in a laboratory setting. There, more accurate information about the effects on repulsion can be measured than it was possible in the online study used.

While our results contribute to interruption research, further studies have to be conducted to investigate the impact of task choice and other (computer) systems, such as mobile phones or tablets.
\bibliographystyle{apa}
\newpage
\bibliography{Draft}

\begin{thebibliography}{}

\bibitem[\protect\astroncite{Addas and
  Pinsonneault}{2015}]{InformationTechnologyInteruption}
Addas, S. and Pinsonneault, A. (2015).
\newblock The many faces of information technology interruptions: a taxonomy
  and preliminary investigation of their performance effects.
\newblock {\em Information System Journals}, 25:231--273.

\bibitem[\protect\astroncite{Anwyl-Irvine
  et~al.}{2020}]{AccuracyOnlineExperiments}
Anwyl-Irvine, A., Dalmaijer, E.~S., Hodges, N., and Evershed, J.~K. (2020).
\newblock Realistic precision and accuracy of online experiment platforms, web
  browsers, and devices.
\newblock {\em Behavior Research Methods}, 53.

\bibitem[\protect\astroncite{Bergmann et~al.}{1981}]{SRTProblemSolvingBehavior}
Bergmann, H., Brinkmann, A., and Koelega, H.~S. (1981).
\newblock {System Response Time and Problem Solving Behavior}.
\newblock {\em Proceedings of the Human Factors and Ergonomics Society Annual
  Meeting}, 25(1):749--753.

\bibitem[\protect\astroncite{Bortz and Schuster}{2010}]{TTest}
Bortz, J. and Schuster, C. (2010).
\newblock {\em {Statistik für Human- und Sozialwissenschaftler}}.
\newblock Springer-Lehrbuch. Springer, Berlin, Heidelberg.

\bibitem[\protect\astroncite{Bridges et~al.}{2020}]{TimingMegaStudy}
Bridges, D., Pitiot, A., MacAskill, M.~R., and Peirce, J.~W. (2020).
\newblock The timing mega-study: comparing a range of experiment generators,
  both lab-based and online.
\newblock {\em Brain, Cognition and Mental Health}.
\newblock {P}eerJ 8:e9414.

\bibitem[\protect\astroncite{Butler}{1983}]{SRTUserResponseTime}
Butler, T.~W. (1983).
\newblock Computer response time and user performance.
\newblock In {\em Proceedings of the SIGCHI Conference on human factors in
  computing systems}, pages 58--62.

\bibitem[\protect\astroncite{Clark and Molnar}{1964}]{HistoryComputerScience}
Clark, W.~A. and Molnar, C.~E. (1964).
\newblock {The LINC: A Description of the laboratory instrument Computer}.
\newblock {\em Annals of the New York Academy of Sciences}, 115(2):653--668.

\bibitem[\protect\astroncite{Colom et~al.}{2000}]{Intelligence}
Colom, R., Juan-Espinosa, M., Abad, F., and Garciá, L.~F. (2000).
\newblock {Negligible Sex Differences in General Intelligence}.
\newblock {\em Intelligence}, 28(1):57--68.

\bibitem[\protect\astroncite{Cowan}{2008}]{MemoryCapacity}
Cowan, N. (2008).
\newblock {\em {Progress in Brain Research}}, volume 169, chapter What are the
  differences between long-term, short-term, and working memory?, pages
  323--338.
\newblock Elsevier.

\bibitem[\protect\astroncite{Crump et~al.}{2013}]{ReportDropouts}
Crump, M. J.~C., McDonnell, J.~V., and Gureckis, T.~M. (2013).
\newblock {Evaluating Amazon’s Mechanical Turk as a Tool for Experimental
  Behavioral Research}.
\newblock {\em PLOS Online}, 8(3):1--18.

\bibitem[\protect\astroncite{Dandurand et~al.}{2008}]{CauseDropout}
Dandurand, F., Shultz, T.~R., and Onishi, K.~H. (2008).
\newblock Comparing online and lab methods in a problem-solving experiment.
\newblock {\em Behavior Research Methods}, 40:428--434.

\bibitem[\protect\astroncite{Davis et~al.}{2015}]{ImageRates}
Davis, J., Hsieh, Y.-H., and Lee, H.-C. (2015).
\newblock Humans perceive flicker artifacts at 500 {H}z.
\newblock {\em Scientific Reports}, 5(1).

\bibitem[\protect\astroncite{de~Leeuw and Motz}{2016}]{JavaScriptPsychophysics}
de~Leeuw, J.~R. and Motz, B.~A. (2016).
\newblock {Psychophysics in a Web browser? {C}omparing response times collected
  with JavaScript and Psychophysics Toolbox in a visual search task}.
\newblock {\em Behavior Research Methods}, 48(1):1--12.

\bibitem[\protect\astroncite{Deng et~al.}{2016}]{AversionGender}
Deng, Y., Chang, L., Yang, M., Huo, M., and Zhou, R. (2016).
\newblock Gender {D}ifferences in {E}motional {R}esponse: {I}nconsistency
  between {E}xperience and {E}xpressivity.
\newblock {\em PLoS ONE}, 11(6).

\bibitem[\protect\astroncite{Dix et~al.}{2004}]{HCIOverall}
Dix, A., Finley, J., Abowd, G.~D., and Beale, R. (2004).
\newblock {\em {Human-Computer Interaction}}.
\newblock Pearson Education, 3 edition.

\bibitem[\protect\astroncite{Faul et~al.}{2009}]{GPower2}
Faul, F., Erdfelder, E., Buchner, A., and Lang, A.-G. (2009).
\newblock {Statistical power analyses using G*Power 3.1: Tests for correlation
  and regression analyses}.
\newblock {\em Behavior Research Methods}, 41:1149--1160.

\bibitem[\protect\astroncite{Faul et~al.}{2007}]{GPower1}
Faul, F., Erdfelder, E., Lang, A.-G., and Buchner, A. (2007).
\newblock {G*Power 3: A flexible statistical power analysis program for the
  social, behavioral, and biomedical sciences}.
\newblock {\em Behavior Research Methods}, 39:175--191.

\bibitem[\protect\astroncite{Fozard et~al.}{1994}]{AgeDifferences}
Fozard, J.~L., Vercruyssen, M., Reynolds, S.~L., Hancock, P.~A., and Quilter,
  R.~E. (1994).
\newblock {Age Differences and Changes in Reaction Time: The Baltimore
  Longitudinal Study of Aging}.
\newblock {\em Journal of Gerontology: PSYCHOLOGICAL SCIENCE}, 49(4):179--189.

\bibitem[\protect\astroncite{Gale et~al.}{2016}]{AgeAndReactionTime}
Gale, C.~R., Harris, A., and Deary, I.~J. (2016).
\newblock {Reaction time and onset of psychological distress: the UK Health and
  Lifestyle Survey}.
\newblock {\em Journal of Epidemiology and Community Health}, 70:813--817.

\bibitem[\protect\astroncite{Galletta et~al.}{2004}]{WebsiteDelays}
Galletta, D., Henry, R., McCoy, S., and Polak, P. (2004).
\newblock {Web Site Delays: How Tolerant are Users?}
\newblock {\em Journal of the Association for Information Systems}, 5(1):1--28.

\bibitem[\protect\astroncite{Grace-Martin}{2020}]{AnovaUnbalanced}
Grace-Martin, K. (2020).
\newblock When {U}nequal {S}ample {S}izes {A}re and {A}re {NOT} a {P}roblem in
  {ANOVA}.

\bibitem[\protect\astroncite{Hilbig}{2016}]{LabVsWeb}
Hilbig, B.~E. (2016).
\newblock {Reaction time effects in lab- versus Web-based research:
  Experimental evidence}.
\newblock {\em 48}, 48:1718--1724.

\bibitem[\protect\astroncite{Holden et~al.}{2020}]{MethodIssuesRT}
Holden, J., Francisco, E., Tommerdahl, A., Lensch, R., Kirsch, B., Zai, L.,
  Pearce, A.~J., Favorov, O.~V., Dennis, R.~G., and Tommerdahl, M. (2020).
\newblock {Methodological Problems With Online Concussion Testing}.
\newblock {\em frontiers in Human Neuroscience}, 14.
\newblock Art. Number: 509091.

\bibitem[\protect\astroncite{Howell}{nD}]{DropoutInternet}
Howell, B. (n.D.).
\newblock Why do participants drop out of online surveys and experiments?

\bibitem[\protect\astroncite{Hultsch et~al.}{2002}]{AgeDifferences2}
Hultsch, D.~F., MacDonald, S. W.~S., and Dixon, R.~A. (2002).
\newblock {Variability in Reaction Time Performance of Younger and Older
  Adults}.
\newblock {\em The Journals of Gerontology}, 57(2):101--115.

\bibitem[\protect\astroncite{Jonides et~al.}{2008}]{MemoryInterference}
Jonides, J., Lewis, R.~L., Nee, D.~E., Lustig, C.~A., Berman, M.~G., and Moore,
  K.~S. (2008).
\newblock {The Mind and Brain of Short-Term Memory}.
\newblock {\em Annual Review of Psychology}, 59:193--224.

\bibitem[\protect\astroncite{Jorm et~al.}{2004}]{CognitiveDifferenceGender2}
Jorm, A.~F., Anstey, K.~J., Christensen, H., and Rodgers, B. (2004).
\newblock Gender differences in cognitive abilities: {T}he mediating role of
  health state and health habits.
\newblock {\em Intelligence}, 32(1):7--23.

\bibitem[\protect\astroncite{Komarov et~al.}{2013}]{UIPerformanceEvaluation}
Komarov, S., Reinecke, K., and Gajos, K.~Z. (2013).
\newblock Crowdsourcing performance evaluations of user interfaces.
\newblock In {\em CHI '13: Proceedings of the SIGCHI Conference on Human
  Factors in Computing Systems}, pages 207--216.

\bibitem[\protect\astroncite{Langner et~al.}{2009}]{MentalFatigue}
Langner, R., Steinborn, M.~B., Chatterjee, A., Sturm, W., and Willmes, K.
  (2009).
\newblock Mental fatigue and temporal preparation in simple reaction-time
  performance.
\newblock {\em Acta Psychologica}, 133(1):64--72.

\bibitem[\protect\astroncite{Matsuno and Budge}{2017}]{Nonbinary}
Matsuno, E. and Budge, S.~L. (2017).
\newblock {Non-binary/Genderqueer Identities: a Critical Review of the
  Literature}.
\newblock {\em Current Sexual Health Reports}, 9(3):116--120.

\bibitem[\protect\astroncite{Miller}{1968}]{RTConversationalTransactions}
Miller, R.~B. (1968).
\newblock Response time in man-computer conversational transactions.
\newblock In {\em AFIPS '68 (Fall, part I)}, volume~1 of {\em AFIPS '68}, pages
  267--277.

\bibitem[\protect\astroncite{Neath et~al.}{2011}]{AccuracyMac}
Neath, I., Earle, A., Hallett, D., and Surprenant, A.~M. (2011).
\newblock {Response time accuracy in Apple Macintosh computers}.
\newblock {\em Behavior Research Methods}, 43.
\newblock Art. Number 353.

\bibitem[\protect\astroncite{Nuerk et~al.}{2004}]{NumberDifferences}
Nuerk, H.-C., Weger, U., and Willmes, K. (2004).
\newblock On the perceptual generality of the unit-decade compatibility effect.
\newblock {\em Experimental Psychology}, 51(1):72--79.

\bibitem[\protect\astroncite{Oberbauer}{2009}]{OverwritingMemory}
Oberbauer, K. (2009).
\newblock Interference between storage and processing in working memory:
  {F}eature overwriting, not similarity-based competition.
\newblock {\em Memory and Cognition}, 37(3):346--357.

\bibitem[\protect\astroncite{Ollesch et~al.}{2006}]{Experimenter}
Ollesch, H., Heineken, E., and Schulte, F.~P. (2006).
\newblock {Physical or Virtual Presence of the Experimenter: Psychological
  Online-Experiments in Different Settings}.
\newblock {\em International Journal of Internet Science}, 1(1):71--81.

\bibitem[\protect\astroncite{Plant}{2016}]{ReplicationFailureAccuracy}
Plant, R.~R. (2016).
\newblock {A reminder on millisecond timing accuracy and potential replication
  failure in computer-based psychology experiments: An open letter}.
\newblock {\em Behavior Research Methods}, 48:408--411.

\bibitem[\protect\astroncite{Reimers and Stewart}{2015}]{RTAccuracy}
Reimers, S. and Stewart, N. (2015).
\newblock {Presentation and response timing accuracy in Adobe Flashand
  HTML5/JavaScript Web experiments}.
\newblock {\em Behavior Research Methods}, 47:309--327.

\bibitem[\protect\astroncite{Reips}{2002}]{DosAndDonts}
Reips, U.-D. (2002).
\newblock {Internet-Based Psychological Experimenting: Five Dos and Five
  Don’ts}.
\newblock {\em Social Science Computer Review}, 20(3):241--249.

\bibitem[\protect\astroncite{Restle}{1970}]{SpeedReaction}
Restle, F. (1970).
\newblock Speed of adding and comparing numbers.
\newblock {\em Journal of Experimental Psychology}, 83(2):274--278.

\bibitem[\protect\astroncite{Rosenthal and Fode}{1963}]{VEE}
Rosenthal, R. and Fode, K.~L. (1963).
\newblock The effect of experimenter bias on the performance of the albino rat.
\newblock {\em Behavioral Science}, 8(3):183--189.

\bibitem[\protect\astroncite{Sanders and Hoogenboom}{1970}]{PracticeReaction}
Sanders, A.~F. and Hoogenboom, W. (1970).
\newblock On the effects of continuous active work on performance.
\newblock {\em Acta Psychologica}, 33:414--431.

\bibitem[\protect\astroncite{Schleifer and Amick}{1989}]{SRTMethodOfPay}
Schleifer, L.~M. and Amick, B.~C. (1989).
\newblock System response time and method of pay: Stress effects in
  computer‐based tasks.
\newblock {\em International journal of human-computer interaction},
  1(1):23--39.

\bibitem[\protect\astroncite{Schmid et~al.}{2020}]{InteractionInterferences}
Schmid, P., Malacria, S., Cockburn, A., and Nancel, M. (2020).
\newblock {Interaction Interferences: Implications of Last-Instant System State
  Changes}.
\newblock In {\em UIST '20: Proceedings of the 33rd Annual ACM Symposium on
  User Interface Software and Technology}, pages 516--528.

\bibitem[\protect\astroncite{Schmidt}{2001}]{WebAnimationMethods}
Schmidt, W.~C. (2001).
\newblock {P}resentation accuracy of {W}eb animation methods.
\newblock {\em Behavior Research Methods, Instruments, \& Computers},
  33:187--200.

\bibitem[\protect\astroncite{Segieth et~al.}{2004}]{PowerAnalysis}
Segieth, C., Ruhleder, M., L.Vogt, and Banzer, W. (2004).
\newblock {Poweranalyse und optimaler Stichprobenumfang – Eine Einführung /
  Power analysis and optimal sample size – An introduction}.
\newblock {\em Deutsche Zeitschrift für Akupunktur}, 47(1):50--51.

\bibitem[\protect\astroncite{Shneiderman}{1984}]{DisplayRate}
Shneiderman, B. (1984).
\newblock Response time and display rate in human performance with computers.
\newblock {\em ACM Computing Surveys}, 16(3):265--285.

\bibitem[\protect\astroncite{Szameitat et~al.}{2009}]{EffectsOfDelay}
Szameitat, A.~J., Rummel, J., Szameitat, D.~P., and Sterr, A. (2009).
\newblock Behavioral and emotional consequences of brief delays in
  human–computer interaction.
\newblock {\em International journal of human-computer studies},
  67(7):561--570.

\bibitem[\protect\astroncite{Teleb and
  Awamleh}{2012}]{CognitiveDifferenceGender}
Teleb, A.~A. and Awamleh, A. A.~A. (2012).
\newblock {Gender Differences in Cognitive Abilites}.
\newblock {\em Current Research in Psychology}, 3.
\newblock Article Number: 41219.

\bibitem[\protect\astroncite{Thomsen et~al.}{2005}]{AversionGender2}
Thomsen, D.~K., Mehlsen, M.~Y., Viidik, A., Sommerlund, B., and Zachariae, R.
  (2005).
\newblock Age and gender differences in negative affect—{I}s there a role for
  emotion regulation?
\newblock {\em Personality and Individual Differences}, 38(8):1935--1946.

\bibitem[\protect\astroncite{Troyer}{2011}]{PositionEffect}
Troyer, A.~K. (2011).
\newblock {\em {Encyclopedia of Clinical Neuropsychology}}, chapter Serial
  Position Effect, page 166.
\newblock Positionseffekt.

\bibitem[\protect\astroncite{Whelan}{2008}]{DataAnalysis}
Whelan, R. (2008).
\newblock Effective analysis of reaction time data.
\newblock {\em The Psychological Record volume}, 58:475--482.

\bibitem[\protect\astroncite{Woods et~al.}{2015}]{LatencySimpleReactionTime}
Woods, D.~L., Wyma, J.~M., {William Yund}, E., Herron, T.~J., and Reed, B.
  (2015).
\newblock Factors influencing the latency of simple reaction time.
\newblock {\em Frontiers in Human Neuroscience}, 9(131):1--12.

\end{thebibliography}
\end{document}